\documentclass[10pt,journal,compsoc]{IEEEtran}

\ifCLASSOPTIONcompsoc
   \usepackage[nocompress]{cite}
\else
    \usepackage{cite}
\fi

\ifCLASSINFOpdf
 
\else
 
\fi
\usepackage{multirow}
\usepackage{graphicx}
\usepackage{xcolor}
\usepackage{tabularx}

\usepackage{authblk}

\begin{document}

\title{Implementation of a noisy hyperlink removal system: A semantic and relatedness approach}


\author{\IEEEauthorblockN{Kazem Taghandiki\IEEEauthorrefmark{1},
Elnaz Rezaei Ehsan\IEEEauthorrefmark{2}}
	\IEEEauthorblockA{
	 \\
		\IEEEauthorrefmark{1}Department of Computer Engineering, \\Technical and Vocational University (TVU), Tehran, Iran
 \\ktaghandiki@tvu.ac.ir
 \\
					\IEEEauthorrefmark{2} Master's degree, industrial engineering,\\ System management and productivity, Iran University of Science and Technology\\  elnazrezaeie110@gmail.com  \\
							}
     
							}




\IEEEtitleabstractindextext{%
\begin{abstract}
\textcolor{black}{As the volume of data on the web grows, the web structure graph, which is a graph representation of the web, continues to evolve. The structure of this graph has gradually shifted from content-based to non-content-based. Furthermore, spam data, such as noisy hyperlinks, in the web structure graph adversely affect the speed and efficiency of information retrieval and link mining algorithms. Previous works in this area have focused on removing noisy hyperlinks using structural and string approaches. However, these approaches may incorrectly remove useful links or be unable to detect noisy hyperlinks in certain circumstances. In this paper, a data collection of hyperlinks is initially constructed using an interactive crawler. The semantic and relatedness structure of the hyperlinks is then studied through semantic web approaches and tools such as the DBpedia ontology. Finally, the removal process of noisy hyperlinks is carried out using a reasoner on the DBpedia ontology. Our experiments demonstrate the accuracy and ability of semantic web technologies to remove noisy hyperlinks. }
\end{abstract}

}

\maketitle

\IEEEdisplaynontitleabstractindextext

\IEEEpeerreviewmaketitle

\section{Introduction}
In recent years, the ability to create any number of web pages, in addition to the extremely large volumes of data created in various fields of technology \cite{1y}, have resulted in a challenging concept known as “Big Data” \cite{a1,a2}. In 2021, nearly 49 billion web pages were indexed by Google and Bing crawlers \cite{10}. Clearly, there is a significant surge in number of web pages on the Internet, which has led to the growth of the web structure graph. As a result, it is exceedingly difficult to navigate and explore the structure of the web, due to spam data such as noisy hyperlinks \cite{1y}. Therefore, the need for a mechanism to eliminate spam hyperlinks is evident. Historically, many information retrieval algorithms used contents of web documents to classify, cluster, and remove spam pages. However, this process-intensive and time-consuming approach was later replaced with algorithms which rely on hyperlink characteristics in the web structure graph instead of document content. A number of algorithms such as PageRank \cite{21} work with these characteristics to reduce the required processing for search engines. However, these link-based algorithms are predicated on the assumption that the links point exactly to pages wanted by the users \cite{8,a3}. Thus, many link mining algorithms, mistakenly, assume that the web structure graph is completely semantic and content-based. However, the graph contains useless spam links, which may both mislead the user and affect the output of the algorithm. Spam hyperlinks allow irrelevant documents to obtain higher ranks than relevant ones. This attempt to boost the rank of a page is mainly made for “business” purposes \cite{8}. A number of studies have been conducted to detect and eliminate spam links from the structure of web. The proposed methods, however, are highly dependent on the string and structural characteristics of hyperlinks \cite{2,8} while ignoring their semantic and relatedness structures. In this paper, we consider the semantic and relatedness structures of the hyperlinks at both the page and site levels. Using semantic web technologies, e.g. ontologies and reasoners, we remove noisy hyperlinks. 
In doing so, initially, a dataset of hyperlinks is created in a separate process. Then, using semantic web technologies such as ontologies and reasoners, the concept of the source page hyperlinks and the concept of the target page are semantically and relationally analyzed. The analysis may be used to determine whether a hyperlink is noisy or useful. 

The proposed system takes the constructed dataset as input; each row of the dataset is composed of the class mapped from the hyperlink context topic of the source page, the class mapped from the topic of the target page, a field indicating the noisy or useful nature of the hyperlink from the user’s perspective, and the domain name of the source page. Thereafter, noisy hyperlinks are detected and the results are compared to those of the user in order to identify the extent to which each semantic or relatedness property in the ontology contributes to a correct detection. Furthermore, among other advantages, it is possible to know which queries lead the user to noisy hyperlinks and which domains contain the greatest number of noisy hyperlinks \cite{a4}. The experiments demonstrate the accuracy, capability, and scalability of semantic web technologies in eliminating noisy hyperlinks. 
The remainder of this paper is organized as follows. In Section 2, a survey of previous works on removal of noisy hyperlinks is given. The implementation of the proposed approach is detailed in Section 3. Section 4 explains the experiments as well as the obtained results. Finally, in Section 5, concluding remarks are presented. 

\section{Related work }\label{ExSurv}
Toward, the implementation of hyperlink removal systems, there are several viable types of research such as, \cite{ a5,a6}. In this regard, Qi et al. \cite{ 17} categorize navigational, advertising, and irrelevant hyperlinks as spam, while other hyperlinks are deemed useful. In their algorithm, a Support Vector Machine (SVN) with two classes, namely “qualified” and “unqualified”, is used to detect and filter noisy hyperlinks, employing a total of six string similarity features. By applying the algorithm to a collection of 2.1 million web pages, 23\% of the hyperlinks are classified as “unqualified”. However, this mechanism does not use semantic or relatedness approaches for removal of hyperlinks. 
Wookey et al. \cite{ 19} design a system called Anchor Woman, wherein noisy hyperlinks are detected using the hyperlink structure of a website and divided into three categories:\\ 
•	Multi-arc loops: chains of hyperlinks which form many cycles in the web graph. \\
•	Multiple arcs:  many hyperlinks that point to the same page. \\
•	Recursive cycles: web pages that contain hyperlinks pointing to themselves.\\
The system works by receiving a web address as input and conducting a breadth-first search of its hyperlinks to identify and eliminate noisy ones. Next, a graphical hierarchical representation of the web structure is generated for the user.
A mechanism to detect noisy hyperlinks at the site level is proposed by Carvalho et al. \cite{ 6} who identify two types of spam relationships: (1) mutual reinforcement, wherein two websites are strongly connected by exchanging site-level hyperlinks and (2) alliance among a chain of strongly connected websites. If the number of hyperlinks between the sites exceeds a threshold, they are considered noisy. The proposed algorithm, which takes a structural approach, is able to remove 16.7 percent of the hyperlinks with a Mean Average Precision of 59.16
Chakrabarti \cite{ 5} proposes a finer grained model of the web, in which pages are represented by their Document Object Models, with the resulted DOM trees being interconnected by regular hyperlinks. The method is able to counter “nepotistic clique attacks”, but needs more input data than our algorithms (which are based exclusively on link analysis). Also, since we specifically target noise removal, we are able to identify different types of hyperlinks anomalies.
Samanta et al. \cite{ 1} use graph-based methods to improve the web structure graph and facilitate user navigation. The paper considers the case of UK university websites. Over six million links are extracted from 110 academic websites to form a dataset. A large portion of undesirable links to images, audio, and video files are removed using TextPipe. The number of web documents, path length, and Strongly Connected Components (SCC) are optimized using this approach. However, the proposed methodology relies merely on the type of hyperlink, without considering semantic or relational approaches. The two steps are as follows: \\
1.	Eliminating advertising and navigational hyperlinks which are commonly located near the top of the page.\\
2.	Eliminating the hyperlinks not covered by association or aggregation relationships.\\
An aggregation relationship refers to a hierarchy relationship between two concepts in which the source concept is broader than the target concept. An association relationship is also known as a horizontal relationship, implying that the source and target concepts share the same parent. In other words, two concepts are horizontally related if and only if they have a common parent. The authors reported a recognition rate of 92.89 percent in the removal of navigational hyperlinks. The aggregation relationship conveys a semantic approach; the association relationship, however, cannot be considered a complete relatedness approach. Even thoughPedersen et al. \cite{ 16} use an association or horizontal relationship to demonstrate relatedness similarity between two concepts, horizontal relationships only indicate a “Part Of” relationship between the two. Nevertheless, other relational properties such as Object Properties in ontologies can also be used to represent relatedness similarity.  
Another algorithm for removing noisy hyperlinks, called the Website Structuring Extracting Algorithm (WSE) was proposed by Oguz \cite{ 12} in 2022. The primary objective of the WSE is to eliminate noisy hyperlinks while retaining semantic ones. However, with regard to semantics, the paper focuses on the path structure of the hyperlinks so that the hierarchy of hyperlinks is maintained. For example, assuming four pages, namely A, B, C, and D, hyperlinks from A to B, from B to C, and from C to D, are semantic hyperlinks whereas hyperlinks in the opposite direction are considered noisy. 
In \cite{ 4} the authors propose to detect nepotistic links using language models. In this method, a link is down-weighted if its source and target page are not related based on their language models. This approach is based on the assumption that pages connected by non-nepotistic links must be sufficiently similar. 
Wu and Davison \cite{ 20} propose a two-step algorithm to identify link farms. The first step generates a seed set based on the intersection of in-link and out-links of web pages. The second step expands the seed set to include pages pointing to many pages within the seed set. The links between these identified spam pages are then re-weighted and a ranking algorithm is applied to the modified link graph.
Previous works tend to focus on page-level hyperlinks; however, modern spam sites usually make use of site-level hyperlinks by generating illegal links to other websites thereby improving their rank in Google’s index. Therefore, it is critical to consider site-level hyperlinks. In this paper, we seek to remove noisy hyperlinks at both page and site levels. Another shortcoming in prior studies pertains to the exclusive application of string or structural approaches. Despite their considerable speed in detecting hyperlink types, these approaches sometimes eliminate useful hyperlinks while being unable to detect noisy hyperlinks at other times. For instance, suppose a page containing the text “Bank” (meaning financial institution) points to a page about “Banks” (meaning shore). Using the string approach, the hyperlink is regarded as useful whereas the semantic approach employs hyperlink text information to identify and eliminate the noisy hyperlinks. This paper aims to apply the semantic web approach and current tools such as the DBpedia ontology to consider the semantic and relational structure of hyperlinks and to remove noisy hyperlinks by activating the DBPedia ontology reasoner. The experiments demonstrate the accuracy and ability of semantic web technologies to eliminate noisy hyperlinks.
In the preceding mechanisms as well as others such as \cite{ 1y,a7,3,a8,9,a9}, existing data collections are not used for information retrieval; rather, noisy hyperlinks are removed through analysis of the web structure graph. Therefore, we need to construct a data collection of hyperlinks. In the following section, the details of this procedure according to information retrieval principles \cite{ 14} are presented.

\section{	Proposed Approch}
The semantic and relatedness system for eliminating noisy hyperlinks involves three general steps as shown in Figure 1

   \begin{figure}
    \centering
    \includegraphics[width=9cm]{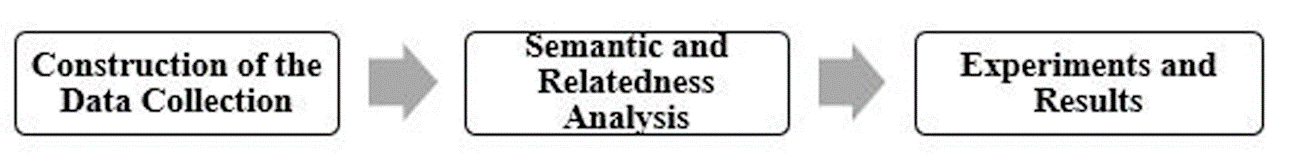}
    \caption{	Constructing the Data Collection}
    \label{fig:life}
\end{figure}

\subsection{Constructing the Data Collection}

The dataset construction step is a distinct process consisting of several steps as shown in Figure 2.

   \begin{figure}
    \centering
    \includegraphics[width=9cm]{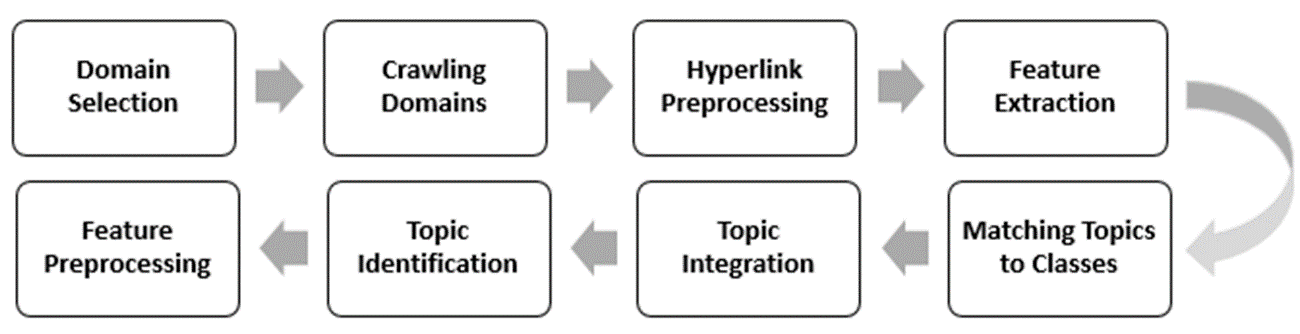}
    \caption{	Stages of constructing the data collection}
    \label{fig:life}
\end{figure}

In constructing the dataset, the user is only involved in domain selection while the other steps are performed independently. 

\subsubsection{Domain Selection}
Internet users use web search engines for a variety of purposes every day, submitting millions of queries to search engines such as Google. Here, deciding which websites to crawl is an important issue, since both noisy and useful hyperlinks are necessary. While the former will be eliminated using semantic and relatedness properties of the ontology, the latter help demonstrates the ability of the proposed approach in detecting useful hyperlinks, using the same properties. The crawled websites must contain content that is popular among Internet users. Using Google Trends, a number of popular topics including news, money, online shopping, new technologies, and celebrities such as actors or athletes were identified. Table 1 shows popular search queries in 2021 according to Google Trends. 

\begin{table}[]
\tiny
\caption{Popular search queries in 2021}
\label{tab:my-table}
\begin{tabular}{|c|c|c|c|}
\hline
\textbf{News}   & \textbf{Celebrities} & \textbf{Sports}            & \textbf{Electronics}    \\ \hline
Afghanistan     & Jenifer   Lawrence   & Real   Madrid CF           & iPhone   13             \\ \hline
AMC   Stock     & Kim   Kardashian     & Chelsea   F.C              & Galaxy   Z Flip4        \\ \hline
COVID   Vaccine & Julie   Gayet        & Paris   Saint-Germain F.C. & Nexus   Summit          \\ \hline
Dogecoin        & Tracy   Morgan       & FC   Barcelona             & Motorola   Moto G Power \\ \hline
\end{tabular}
\end{table}

By learning about popular topics among Internet users, organizations and individuals can take two distinct approaches in designing web pages.\\
1.	Creating websites that are weakly focused on the topic but use background hyperlinks to conduct highly profitable business activities such as directing users to online stores or pornography websites. Such websites contain noisy hyperlinks which have no regard for web user needs.\\
2.	Creating websites with useful content on a particular topic to provide users with appropriate information. Hyperlinks in these websites are rarely considered spam. These websites contain useful hyperlinks that are in line with user needs.
The main idea behind this domain selection approach is to crawl the domains that are retrieved by search engines in response to frequent queries. The retrieved domains fall into one of two general categories: (1) those having noisy hyperlinks for illegal business purposes and (2) those with legal objectives that provide useful hyperlinks to help users achieve their goals. The ability to distinguish between these categories shows the strength of the proposed method in maintaining useful hyperlinks while removing noisy ones.

\subsubsection{Crawling Domains}
In this stage, the user enters each topic from the previous stage into the Google search engine and randomly selects a subset of the returned websites. Then website addresses are given as input to an interactive crawler which is developed using libraries of the Java programming language. Next, the crawler begins exploring the domain; and finally, a list of the links in the domain is obtained, as shown in Figure 3.

   \begin{figure}
    \centering
    \includegraphics[width=9cm]{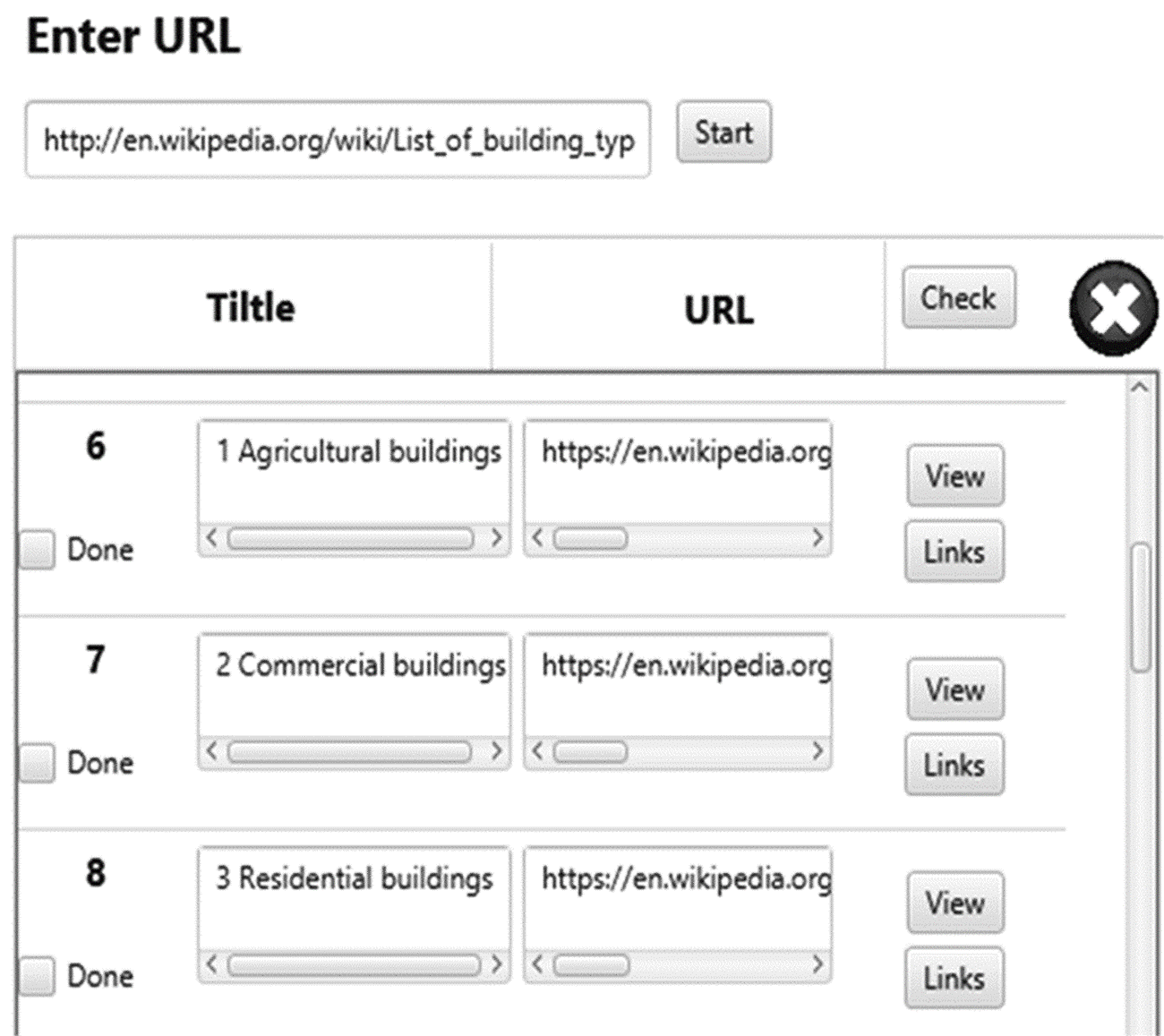}
    \caption{	Links extracted by the crawler}
    \label{fig:life}
\end{figure}

Here, a total of 114 useful and unuseful domains related to the topics were crawled, all of which are in English.

\subsubsection{	Hyperlink Preprocessing }
A large portion of the extracted hyperlinks, e.g. repetitive links or those pointing to audio or video files, are inappropriate for the purpose of this paper. Hence, they are removed according to the operations proposed by [1] (Section 2). The status of the hyperlinks subsequent to preprocessing can be seen in Figure 4.

   \begin{figure}
    \centering
    \includegraphics[width=9cm]{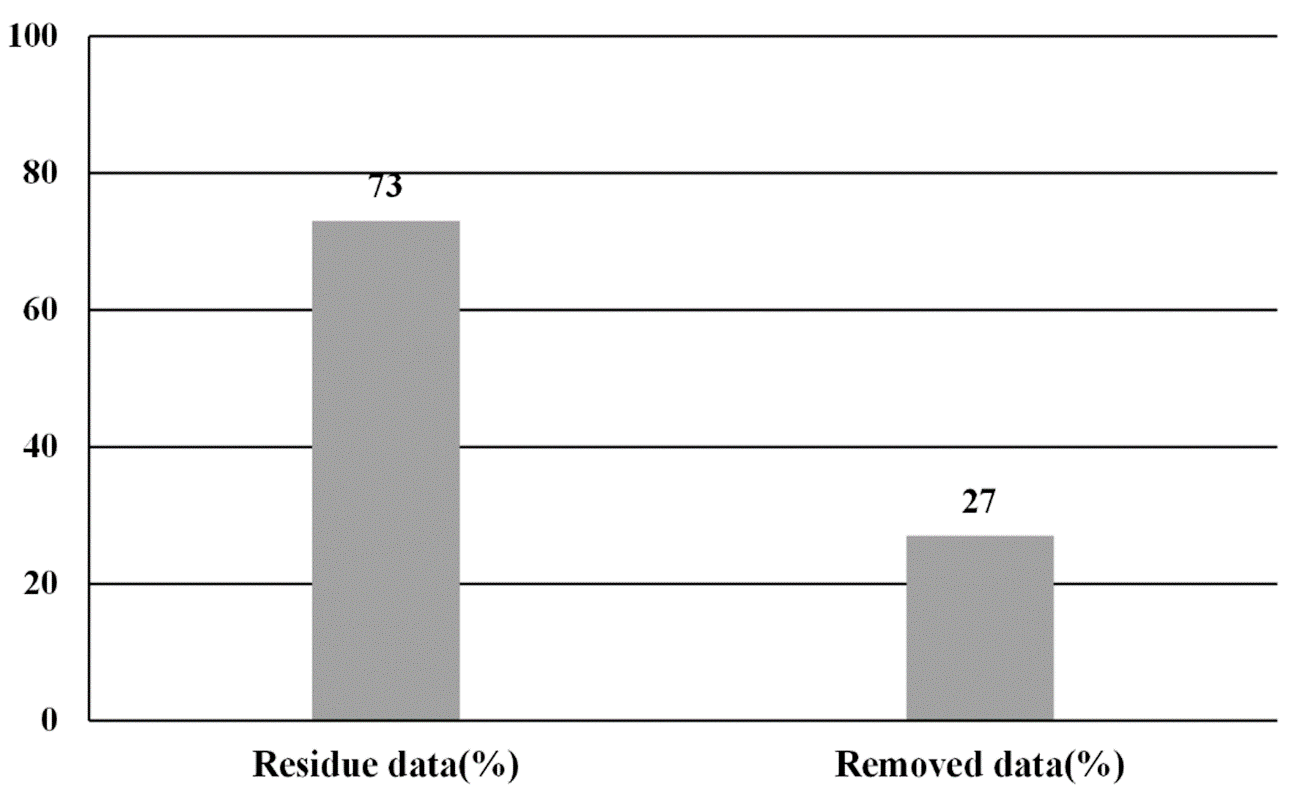}
    \caption{	Status of the hyperlinks subsequent to preprocessing}
    \label{fig:life}
\end{figure}

As shown, following the preprocessing operations, 27 percent of the hyperlinks are removed, reducing the number of crawled hyperlinks from 2665 to 1946. 

Note that a large number of the extracted domains included many video, image, and audio hyperlinks, which are not compatible with the proposed approach. In some cases, the links occurred several times in a page. Therefore, 27 percent of the links (719 links) were eliminated. As mentioned, the proposed method is only compatible with text hyperlinks making the subsequent steps (i.e. those concerning dataset construction) only applicable to text hyperlinks. Thus, incompatible hyperlinks need to be removed in the preprocessing step.

\subsubsection{	Feature Extraction }
In order to ensure that the topic of the hyperlink’s surrounding text as well as that of its target page are detected with sufficient accuracy, a number of features must be extracted from web pages. However, one needs to determine which features are used more frequently for designing web pages; this in turn helps identify the key features to be extracted. The frequencies of five key features in 5000 pages are presented in Figure 5. As shown, “Keyword Metadata”, “Title Tag”, and “First-Level Heading Tag”are most common in web pages.

   \begin{figure}
    \centering
    \includegraphics[width=9cm]{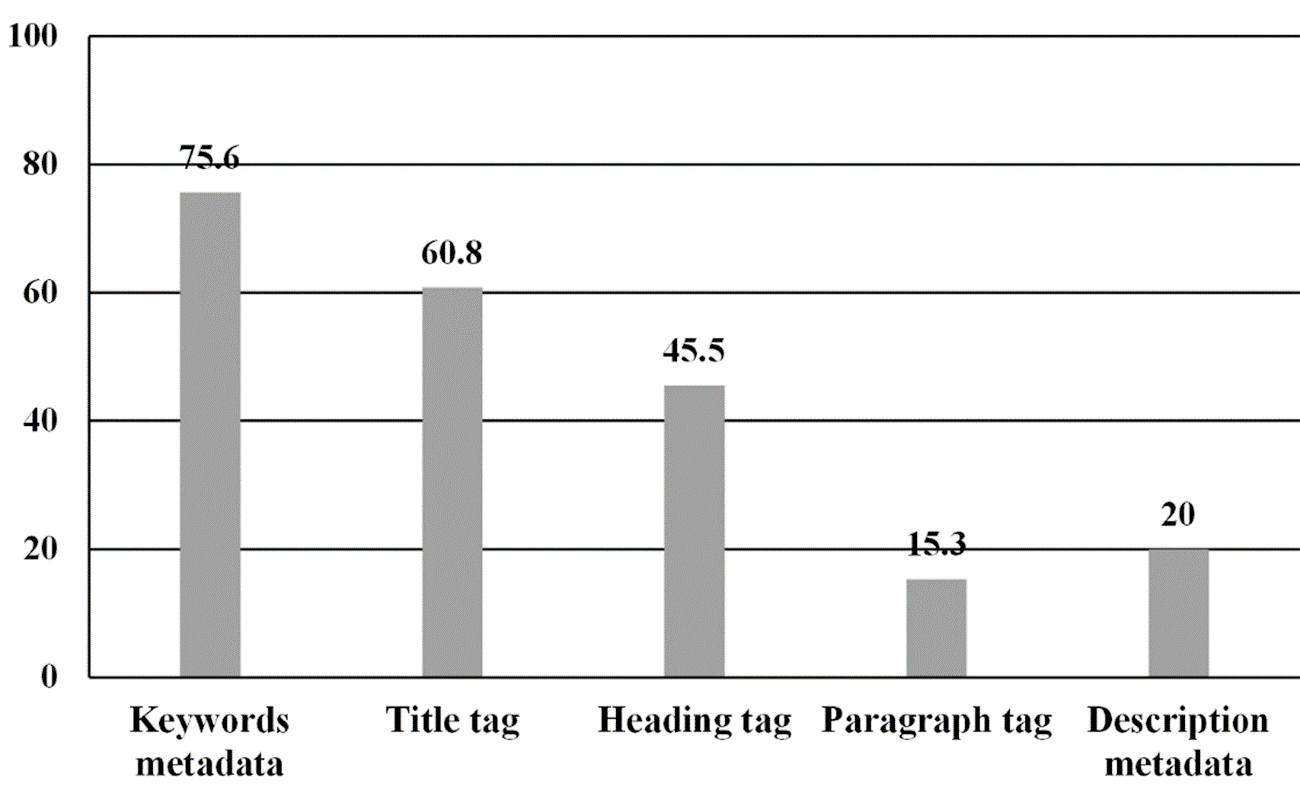}
    \caption{	Top five most commonly used features in web pages}
    \label{fig:life}
\end{figure}

Therefore, three features of the page i.e. title tag, keyword metadata and first-level heading tag are used to determine the topic of the target page. In addition to the first two features, the text of the hyperlink as well as theparagraphcontaining it are utilized to extract the topic context of the hyperlink from the source page. Hyperlink text and its surrounding paragraph act as features that form the context of the hyperlink. Tables 2 and 3 present examples of extracted features for determining the topic of the target page and the context of hyperlink text, respectively.

\begin{table}[]
\caption{Extracted features to detect target page topic (from www.Filehippo.ocm)}
\centering
\tiny
\label{tab:my-table}
\begin{tabular}{|c|c|}
\hline
\textbf{Page Title} & \textbf{Download free Software}                                                         \\ \hline
Keyword Metadata    & \begin{tabular}[c]{@{}c@{}}download software freeware \\ shareware program\end{tabular} \\ \hline
First-level Heading & Software                                                                                \\ \hline
\end{tabular}
\end{table}

\begin{table}[]
\centering
\caption{Extracted features to detect hyperlink text topic context (from www.Filehippo.com)}
\tiny
\label{tab:my-table}
\begin{tabular}{|c|c|}
\hline
\textbf{Page Title} & \textbf{Download free Software}                \\ \hline
Keyword Metadata    & download software freeware   shareware program \\ \hline
Hyperlinktext       & New Software                                   \\ \hline
Hyperlink paragraph & The Latest Versions of the New   Software      \\ \hline
\end{tabular}
\end{table}

These features accelerate the analysis and topic detection procedures in the following stages. In contrast to our work, several studies \cite{a10,7,11,a11, 13,18} extract the topic based on the entire content of the documents. 
Most web designers make use of Web 2.0 techniques such as HTML to create web documents. The markup language is less semantically capable compared to its Web 3.0 counterparts such as RDF, XML, and OWL \cite{a11}. Thus, given the popularity of HTML attributes in designing web documents, the language appears to be the best choice for feature selection in Web 2.0. However, by taking advantage of Web 3.0 techniques, it becomes quite easy to semantically extract features; this in turn plays an important role in the proposed approach. Nevertheless, this is beyond the scope of this paper and remains a recommended direction for future works.

\subsubsection{ Feature Preprocessing }
The quality of the features extracted in the previous step must be improved so that the semantic and relatedness system is able to perform the topic identification process with higher accuracy and lower error rate. In this paper, this is achieved by using typical text mining preprocessing techniques such as stop words, token normalization, case folding, and stemming \cite{14}: \\
1.	Stop words refer to words which occur frequently and are of little use in finding specific information. Examples include articles and prepositions such as “the”, “and”, “or”, etc. Removing these words accelerates the topic detection step. \\
2.	Token normalization is a standardization process that aims to use a single form for each word. For instance, consider “anti discriminatory” and “nondiscriminatory”; subsequent to the normalization process, both forms are mapped onto “anti-discriminatory”. \\
3.	Case folding is a well-known word normalization process wherein capitalized letters are converted to their lower-case equivalents. In fact, case folding may be regarded as a type of token normalization. \\
4.	Words are often used in different forms depending on grammatical rules; for example, organize, organize, and organize. Furthermore, different forms of a word may have nearly similar meanings e.g. democracy and democratic. By removing the endings of the words, the Stemming process aims to obtain a common root for different forms of a word. \\
In this study, preprocessing is performed via MALLET as well as the text mining library in Python.

\subsubsection{  Topic Identification}
The purpose of this stage is to determine the topic of hyperlink text and the target page using the previously extracted features. In this stage, a supervised process with three operations occurs, as depicted in Figure 6. All operations are conducted using MALLET. In the following subsections, pertinent details are provided. 
  \begin{figure}
    \centering
    \includegraphics[width=9cm]{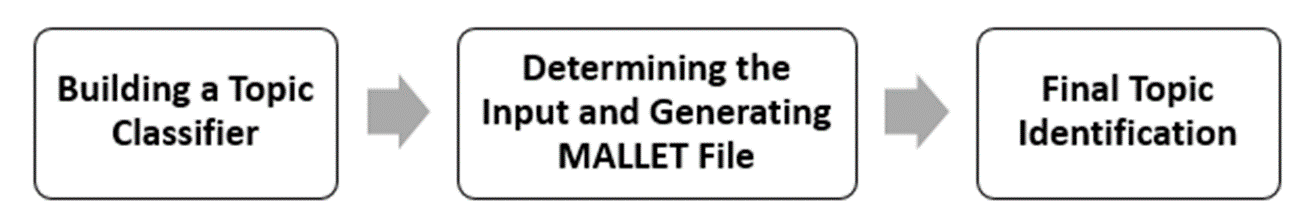}
    \caption{	Supervised topic detection process}
    \label{fig:life}
\end{figure}

1. Building a Topic Classifier \\
In order to build a topic classifier, an initial set of high-quality features is required. Thus, we decided to construct a dataset of features by extracting the three most common key features (i.e. keyword metadata, title tag, and first-level heading tag) of 5000 web pages. The topic classifier is built upon this dataset. Furthermore, 80 percent of the data were used for training purposes while the remaining 20 percent were used for testing the accuracy of the classifier. 
In this paper, topic classification was conducted using four different methods, namely Naïve Bayes, C4.5, Decision Tree, and Max Entropy with 10 cross-validations \cite{a12,cc}, to obtain high efficiency. The algorithms were compared to find the best method for classification. According to Figure 7, Max Entropy yields the highest accuracy on the training data. Therefore, the algorithm is used to create the topic classifier. 
  \begin{figure}
    \centering
    \includegraphics[width=9cm]{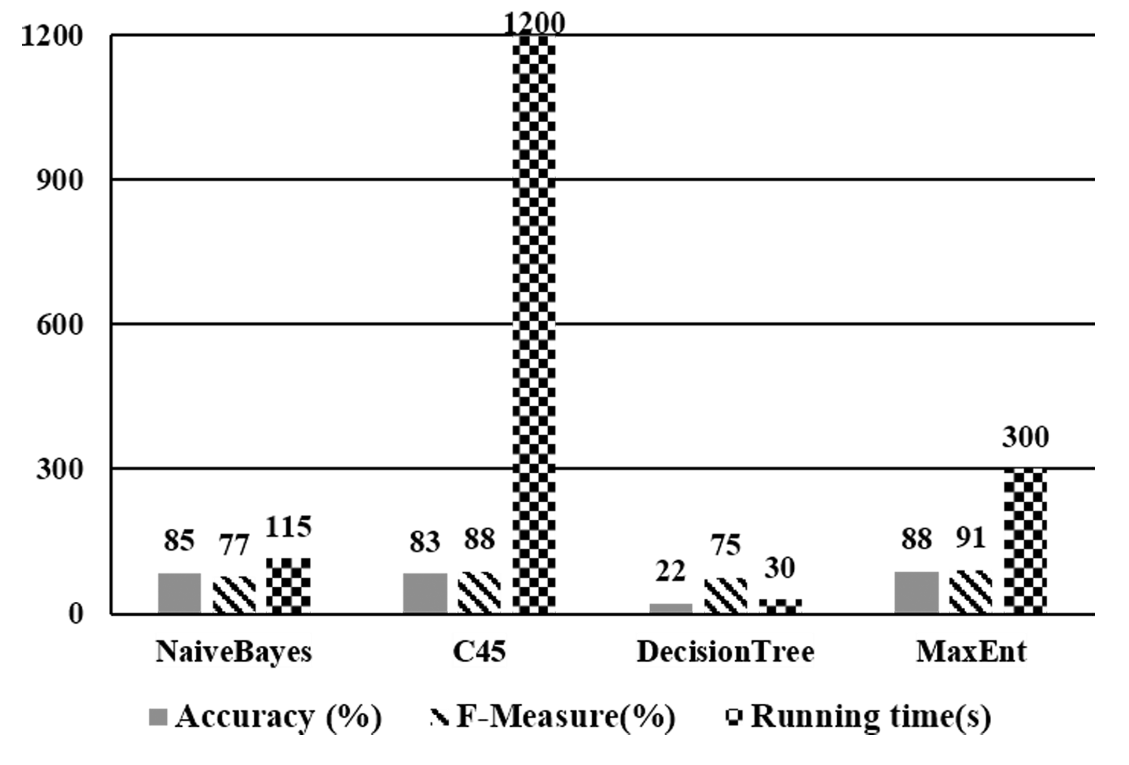}
    \caption{	Performance comparison of Naïve Bayes, Decision Tree, C4.5, and Max Entropy in terms of accuracy and execution time on the features dataset}
    \label{fig:life}
\end{figure}

2. Determining the Input and Generating MALLET File \\
This operation involves providing the extracted features from hyperlink text context and target page as input to MALLET in order to identify the topic of each one. This is carried out using the command-line interface. As a result, two output files are obtained, which are also known as feature vectors. The vectors are numerical representations of the input values that enable faster analysis operations \cite{a13}. The files serve as input to the next operation. 

3. Final Topic Identification \\
Here, the MALLET output files from the previous operation are obtained and the topic classifier identifies the topic. Moreover, the ensuing model (i.e. output) can be used to infer the topic of new input data in the subsequent steps. 
Upon completion, the topic classifier is able to successfully identify the topic of approximately 81.8 percent of extracted features. The remaining 18.2 percent of the features may remain unclassified for one of the following reasons: \\
1.	As a result of poor design, the feature extraction step may be unable to extract appropriate features for identifying the topic of the hyperlink text and the target page. This situation precludes topic assignment. \\
2.	The features used to construct the topic classifier may be completely distinct from those extracted from a new page. Consequently, the page is not assigned a topic. 
The output of the step includes four separate text files. Table 4 presents a portion of each file.\\

\begin{table}[]
\caption{Sample output obtained after the final topic identification step}
\centering
\tiny
\label{tab:my-table}
\begin{tabular}{|c|c|c|c|}
\hline
\textbf{File 4}                                                                                                                            & \textbf{File 3}                                                                   & \textbf{File 2}                                                                                       & \textbf{File 1}                                                                                  \\ \hline
\begin{tabular}[c]{@{}c@{}}Nationalgeographic\\    \\ Nationalgeographic\\    \\ Dairyfoods\\    \\ Songsmp3\\    \\ Songsmp3\end{tabular} & \begin{tabular}[c]{@{}c@{}}0\\    \\ 0\\    \\ 1\\    \\ 1\\    \\ 1\end{tabular} & \begin{tabular}[c]{@{}c@{}}Fish\\    \\ Mammal\\    \\ Canvas\\    \\ Movie\\    \\ Game\end{tabular} & \begin{tabular}[c]{@{}c@{}}Bird\\    \\ Bird\\    \\ Car\\    \\ Song\\    \\ Music\end{tabular} \\ \hline
\end{tabular}
\end{table}

\subsubsection{ Topic Integration }
The topics from the previous stage, located in several text files, are combined and integrated to create a single well-formed input for the matching stage. An example of the output file is shown in Figure 8.

  \begin{figure}
    \centering
    \includegraphics[width=9cm]{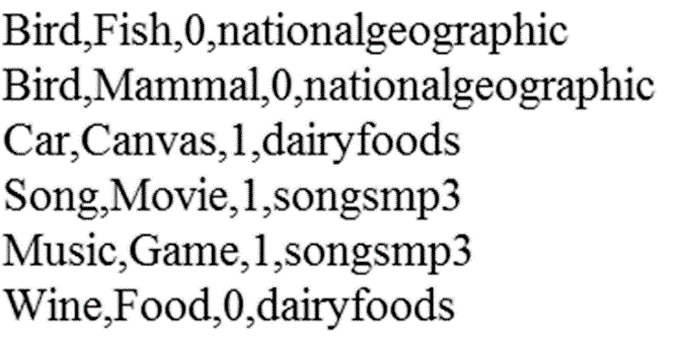}
    \caption{	Topic integration output file}
    \label{fig:life}
\end{figure}

From left to right, the fields in the file represent the context topic of the hyperlink in the source page, topic of the target page, type of the link (i.e. noisy or useful) as perceived by the user, and the domain of the source page. In effect, a topic dataset is constructed, in this stage.

\subsubsection{  Matching Topics to Classes }
The purpose of this stage is to match the topics from the previous stage to classes of the DBpedia ontology. The matching operations are conducted using the WS4J library and the Terminological Search Algorithm (Semantic Search) \cite{a14}. A comparison of the two algorithms using 400 randomly selected topics and five criteria, including accuracy, is shown in Figure 9.

  \begin{figure}
    \centering
    \includegraphics[width=9cm]{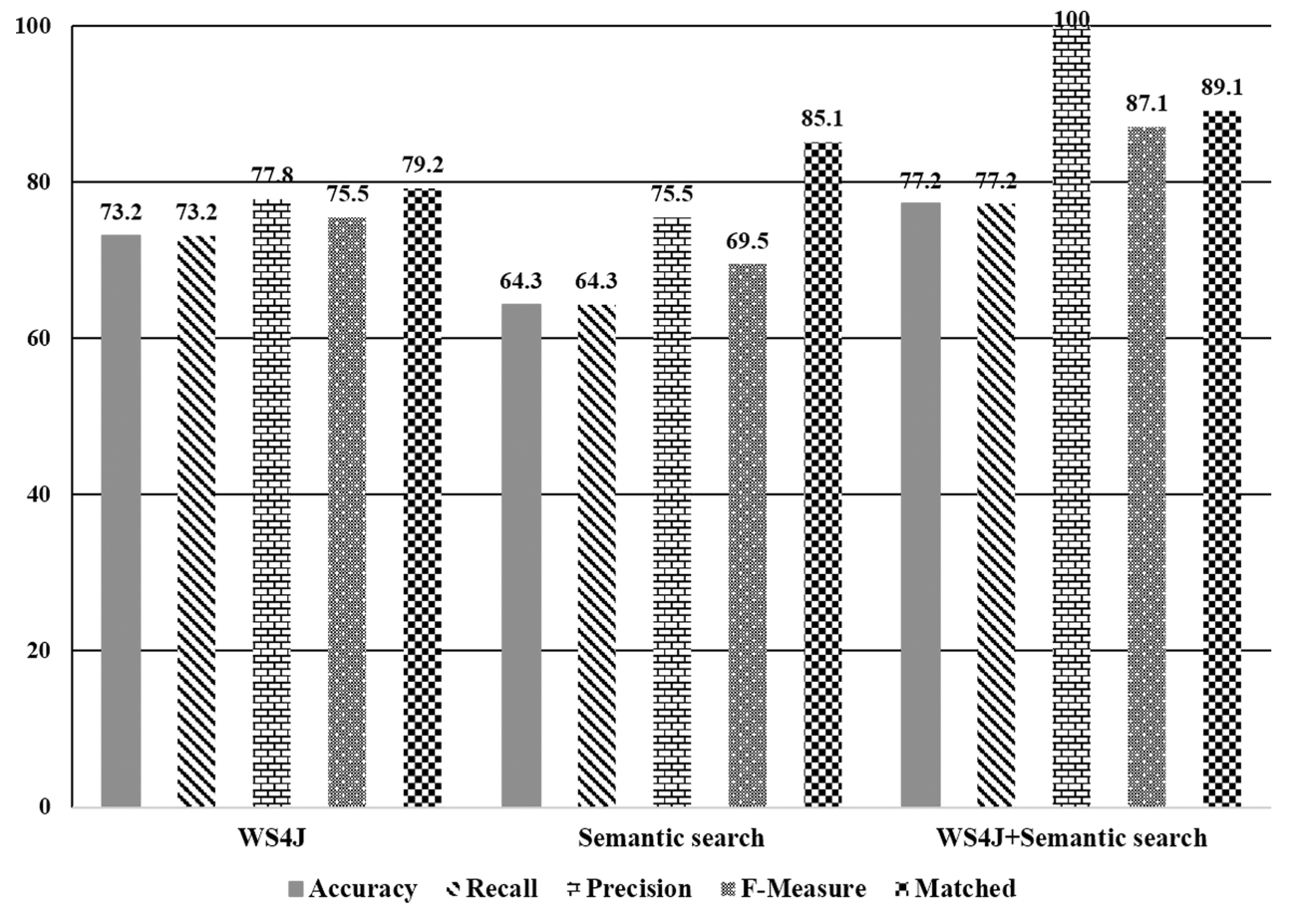}
    \caption{	Comparison of WS4J, Semantic Search, and WS4J+Semantic Search}
    \label{fig:life}
\end{figure}

According to the results, WS4J and Semantic Search may complement each other, which justifies examining their simultaneous application. As shown in Figure 9, the combination of WS4J and Semantic Search outperforms every single algorithm. Furthermore, both algorithms support the required semantic and string similarity. We thus use WS4J+Semantic Search for matching purposes. Examples of the final matches are presented in Figure 10.

  \begin{figure}
    \centering
    \includegraphics[width=9cm]{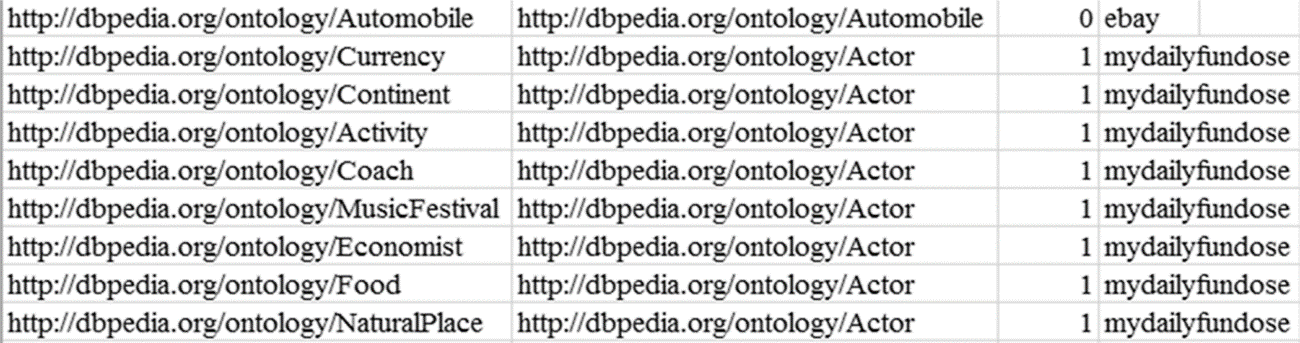}
    \caption{Examples of matches}
    \label{fig:life}
\end{figure}

Once again, the fields represent the class matched to the topic context of the hyperlink in the source page, the topic of the matched class in the target page, the type of hyperlink, and the domain name, respectively. In this step, using the mapping subsystem, nearly 87.3 percent of topics are mapped to those of the DBpedia ontology. In fact, in this step, a final dataset is created which can be referred to as the conceptual hyperlink dataset.

\subsection{Semantic and Relatedness Analysis }
In this stage, the final data collection is given as input to the system, which contains two concepts or ontology classes: the context of the hyperlink text and the target page. The system then examines the relatedness and semantic properties of the input using the added knowledge from the DBpedia ontology. In this way, it is possible to determine whether the hyperlink is noisy or useful. 
The reasoner is the most critical component in the semantic and relatedness analysis step. It is a piece of software which works on one or more conceptual datasets created using ontologies. The reasoner aims to extract logical results from extant facts in the ontology. The proposed system achieves this task using Pellet.
By acting on the primary knowledge from the DBpedia ontology, the Pellet reasoner obtains added knowledge from the properties, relations, and classes of the ontology. Next, a row of the data collection, representing a hyperlink in a web page, is given in the form of two concepts i.e. hyperlink text context and target page. Semantic and relatedness similarities of the two are determined using the properties and relations that are inferred from the DBpedia ontology. In the remainder of this paper, we use the terms source concept and target concept to refer to the concepts of hyperlink text and target page, respectively.\\
A hyperlink is considered useful, by the reasoner, if at least one of the following properties is satisfied: \\
1. “Equivalent Class”: The source and target concepts are equivalent. 
2. “Subclass Of” and “Has Superclass”: The source concept is a sub/superclass of the target concept. \\
Properties (1) and (2) are known as semantic properties, which represent semantic similarity between the two concepts. For instance, the concepts of “Woman” and “Person” are semantically similar since the former is a subclass of the latter.\\ 
3. “Object Property”: The source and target concepts are related through an object property. This is known as a relatedness property and represents relatedness similarity. As an example, the concepts of “Monkey” and “Banana” have relatedness similarity through several relations such as “Liking” or “Eating” (i.e. “The monkey likes bananas” or “The money eats bananas”). 
A number of current noisy hyperlink removal methods merely focus on the first two properties while ignoring Object Properties. As a result, they cannot detect relatedness properties between the source and target concepts which may result in falsely removing a useful hyperlink.

If none of these conditions hold, then the source and target concepts have no semantic or relatedness similarity; thus, the hyperlink is considered noisy. Put differently, instead of taking the user to his/her intended page, the hyperlink on the page leads to an unexpected and completely irrelevant page. Thus, at the end of this step, a conceptual dataset of hyperlinks is created, whose noisy or useful nature is determined by the reasoner. In Figure 11, several records constructed during the analysis stage are given. 

  \begin{figure}
    \centering
    \includegraphics[width=9cm]{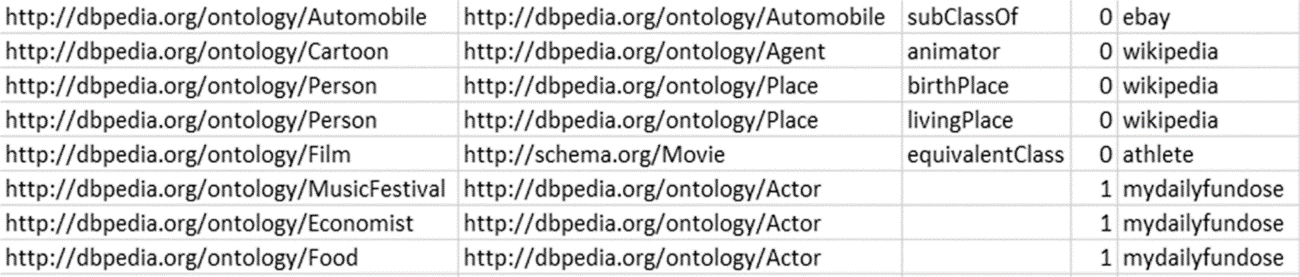}
    \caption{Records constructed during the analysis stage}
    \label{fig:life}
\end{figure}

From left to right, the fields denote Subject, Object, inferred relational and semantic property for the relation between the first two fields, type of hyperlink perceived by the reasoner, and the domain of the source page, respectively. 

\subsection{Experiment and Results }

An important and useful tool for evaluating the proposed approach a confusion matrix which involves two types of labelling: (1) system labelling during semantic and relatedness analysis and (2) expert user labelling while creating the dataset. Each element of the matrix may be one of the following: True Positive (TP), True Negative (TN), False Positive (FP), and False Negative (FN).
Table 5 shows the confusion matrix of the system, which is generated based on user opinions provided during the data collection stage and inferences by the reasoner during the semantic and relatedness analysis.

\begin{table}[]
\centering
\tiny
\caption{ Confusion matrix of the semantic and relatedness system}
\label{tab:my-table}
\begin{tabular}{|ccccc|}
\hline
\multicolumn{5}{|c|}{\textbf{Inferred label}}                                                                                                                                                                                   \\ \hline
\multicolumn{1}{|c|}{\multirow{4}{*}{\textbf{\begin{tabular}[c]{@{}c@{}}Actual label \\ provided by users\end{tabular}}}} & \multicolumn{1}{c|}{}      & \multicolumn{1}{c|}{Yes}       & \multicolumn{1}{c|}{NO}       & Total \\ \cline{2-5} 
\multicolumn{1}{|c|}{}                                                                                                    & \multicolumn{1}{c|}{Yes}   & \multicolumn{1}{c|}{1145 (TP)} & \multicolumn{1}{c|}{232 (FN)} & 1377  \\ \cline{2-5} 
\multicolumn{1}{|c|}{}                                                                                                    & \multicolumn{1}{c|}{NO}    & \multicolumn{1}{c|}{50 (FP)}   & \multicolumn{1}{c|}{519 (TN)} & 569   \\ \cline{2-5} 
\multicolumn{1}{|c|}{}                                                                                                    & \multicolumn{1}{c|}{Total} & \multicolumn{1}{c|}{1195}      & \multicolumn{1}{c|}{751}      & 1946  \\ \hline
\end{tabular}
\end{table}

•	TP represents the number of hyperlinks which are considered useful by both the user and the proposed system.\\
•	FP represents the number of hyperlinks which are considered noisy by the user and useful by the proposed system.\\
•	TN represents the number of hyperlinks which are considered noisy by both the user and the proposed system.\\
•	FN represents the number of hyperlinks which are considered useful by the user and noisy by the proposed system.\\
Values of six commonly used performance measures in information retrieval systems are visualized in Figure 12. As illustrated, the proposed approach achieves high levels of accuracy and precision, while maintaining error rate sufficiently low.

  \begin{figure}
    \centering
    \includegraphics[width=9cm]{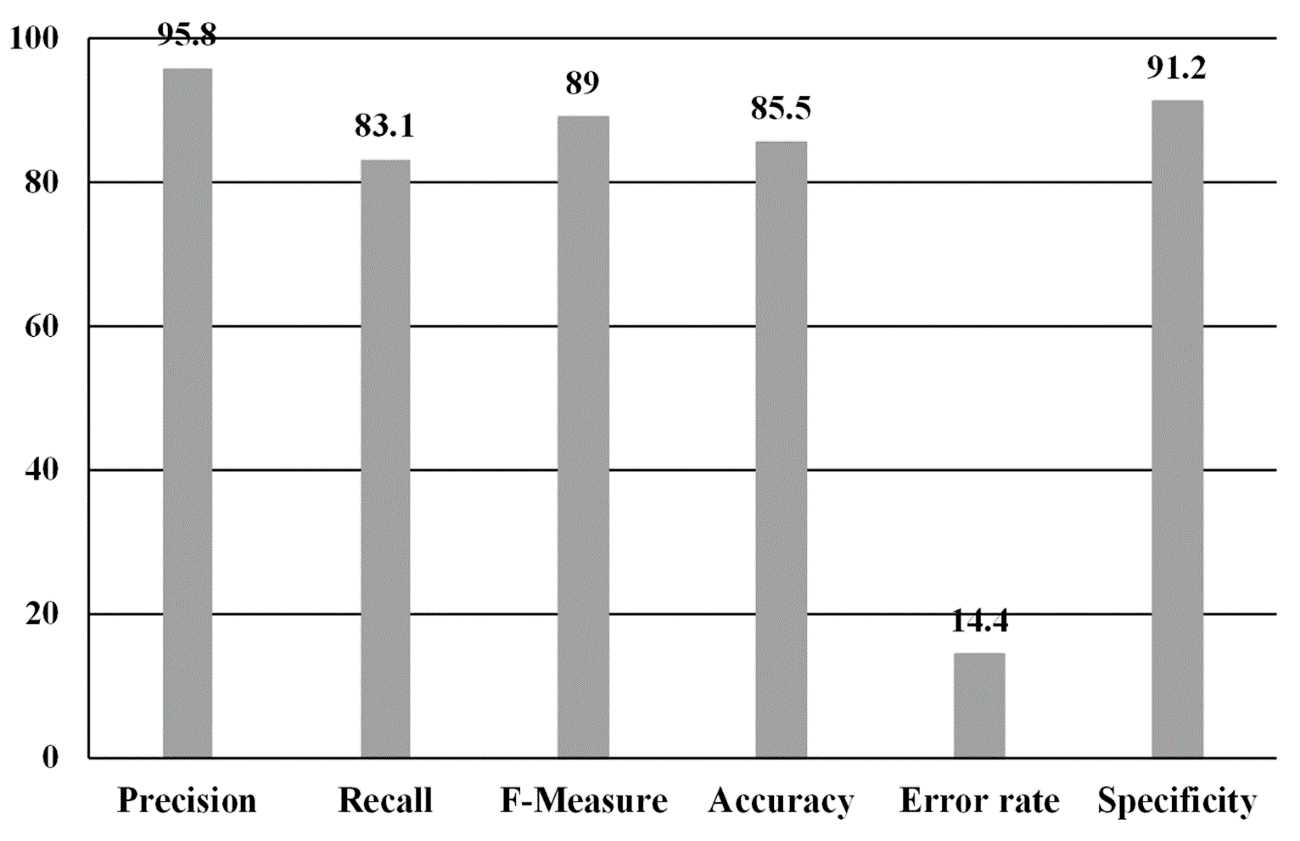}
    \caption{Performance measures of the proposed approach}
    \label{fig:life}
\end{figure}

The extent to which the reasoner uses various semantic and relatedness properties in theDBpedia ontology to represent semantic and relatedness similarities between the Subject and the Object is shown in Figure 13.
  \begin{figure}
    \centering
    \includegraphics[width=9cm]{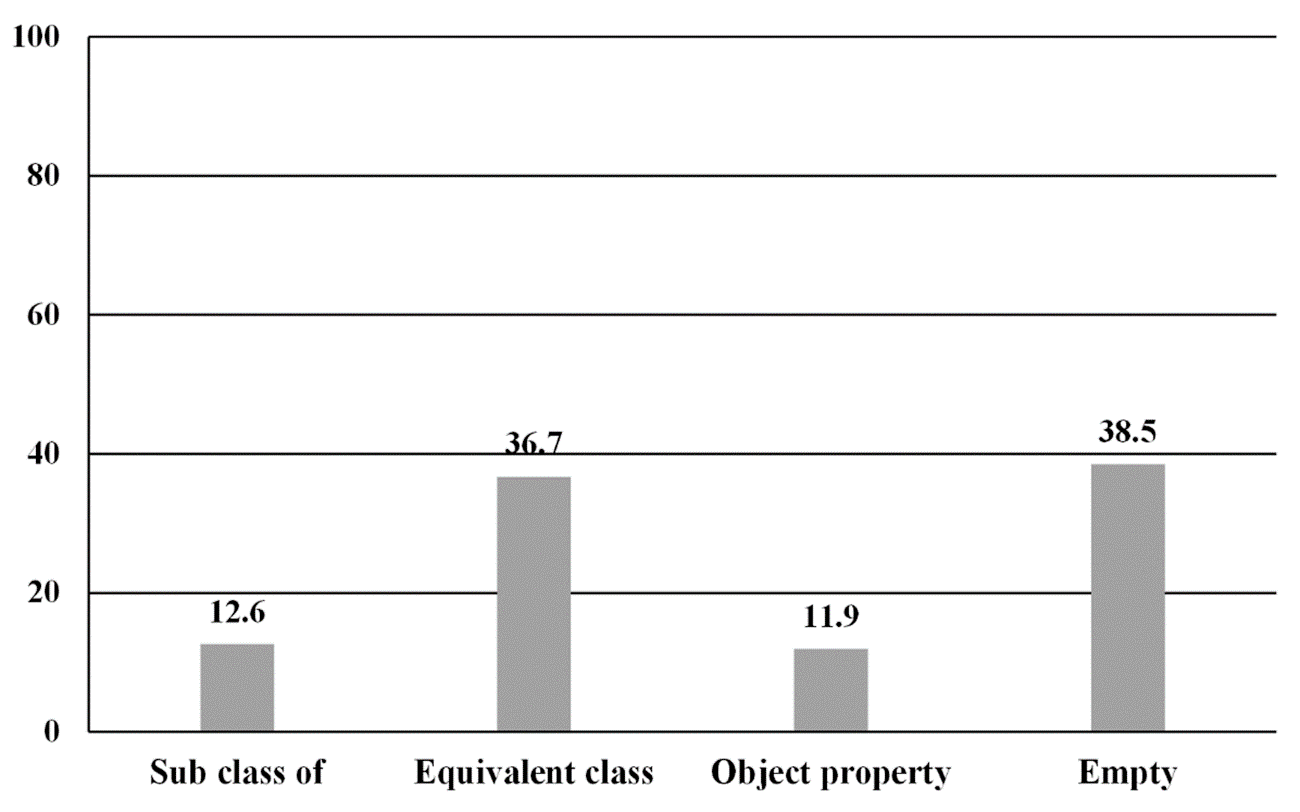}
    \caption{Percentages of properties from the DBpedia ontology used by the reasoner}
    \label{fig:life}
\end{figure}

From Figure 13, one can see that the reasoner uses the “Subclass Of” property in 12.6 percent of the cases to relate the source hyperlink concept to the target page concept. The corresponding value for “Equivalent Class” is 36.7 percent. These properties indicate semantic similarity between the source and target concepts. Furthermore, in 11.9 percent of the cases, the reasoner uses “Object Property”, which represents relatedness similarity. For the remaining 38.5 percent, no semantic or relatedness property is found using the DBpedia ontology. An overview of the results obtained through semantic and relatedness analysis is presented in Table 6.\\

\begin{table}[]
\centering
\tiny
\caption{Semantic and relatedness analysis results}
\label{tab:my-table}
\begin{tabular}{|ccc|}
\hline
\multicolumn{2}{|c|}{\textbf{Number of crawled links}} & \textbf{Number of links in the dataset} \\ \hline
\multicolumn{2}{|c|}{2665}                             & 1946                                    \\ \hline
\multicolumn{2}{|c|}{\textbf{Number of domains}}       & \textbf{Ontology classes}               \\ \hline
\multicolumn{2}{|c|}{114}                              & 312   (38\%)                            \\ \hline
\multicolumn{3}{|c|}{\textbf{Useful domains}}                                                    \\ \hline
\multicolumn{3}{|c|}{Wikipedia,   bbc, Facebook, YouTube, eBay}                                  \\ \hline
\multicolumn{3}{|c|}{\textbf{Noisy domains}}                                                     \\ \hline
\multicolumn{3}{|c|}{mydailyfundose, Google,   fullmoviesfreedownload, hollywoodlife, wikipedia} \\ \hline
\multicolumn{1}{|c|}{Useful domains}         & \multicolumn{2}{c|}{Noisy domains}                \\ \hline
\multicolumn{1}{|c|}{61.4\%}                 & \multicolumn{2}{c|}{38.55\%}                      \\ \hline
\multicolumn{3}{|c|}{\textbf{Target concepts of a noisy link}}                                   \\ \hline
\multicolumn{3}{|c|}{Shopping   Mall, Currency, Film, Actor, Drug}                               \\ \hline
\end{tabular}
\end{table}
The feature extraction step obtains and stores a small set of information pertaining to the hyperlink domain (href values of the <a> tag). Thus, after encountering the hyperlink on the source page, the user determines whether the hyperlink is noisy. The objective is to compare our results with the users’ opinions. 
To better understand the semantic and relational analysis results in Table 6, a comparison with the user's perspective is also conducted. Table 7 provides an overview of the user's perspective identifying the useful or noisy nature of the hyperlinks.

\begin{table}[]
\centering
\tiny
\caption{Analysis of the data collection from the perspective of the user}
\label{tab:my-table}
\begin{tabular}{|ccc|}
\hline
\multicolumn{2}{|c|}{\textbf{Number of crawled links}} & \textbf{Number of links in the dataset} \\ \hline
\multicolumn{2}{|c|}{2665}                             & 1946                                    \\ \hline
\multicolumn{3}{|c|}{\textbf{Number of domains}}                                                 \\ \hline
\multicolumn{3}{|c|}{114}                                                                        \\ \hline
\multicolumn{3}{|c|}{\textbf{Useful domains}}                                                    \\ \hline
\multicolumn{3}{|c|}{Wikipedia,   bbc, Facebook, YouTube, eBay}                                  \\ \hline
\multicolumn{3}{|c|}{\textbf{Noisy domains}}                                                     \\ \hline
\multicolumn{3}{|c|}{mydailyfundose,   Google, fullmoviesfreedownload, hollywoodlife, songmp3}   \\ \hline
\multicolumn{1}{|c|}{Useful domains}         & \multicolumn{2}{c|}{Noisy domains}                \\ \hline
\multicolumn{1}{|c|}{70.75}                  & \multicolumn{2}{c|}{29.2\%}                       \\ \hline
\multicolumn{3}{|c|}{\textbf{Target concepts of a noisy link}}                                   \\ \hline
\multicolumn{3}{|c|}{Shopping   Mall, Currency, Film, Actor, Model}                              \\ \hline
\end{tabular}
\end{table}
A comparison of Tables 6 and 7 testifies to the ability of the semantic and relatedness approach. The percentages of the hyperlink types from both perspectives are visualized in Figure 14. Looking at the figure, it is evident that the proposed approach is able to distinguish between noisy and useful hyperlinks just as accurately as an expert user.
   \begin{figure}
    \centering
    \includegraphics[width=9cm]{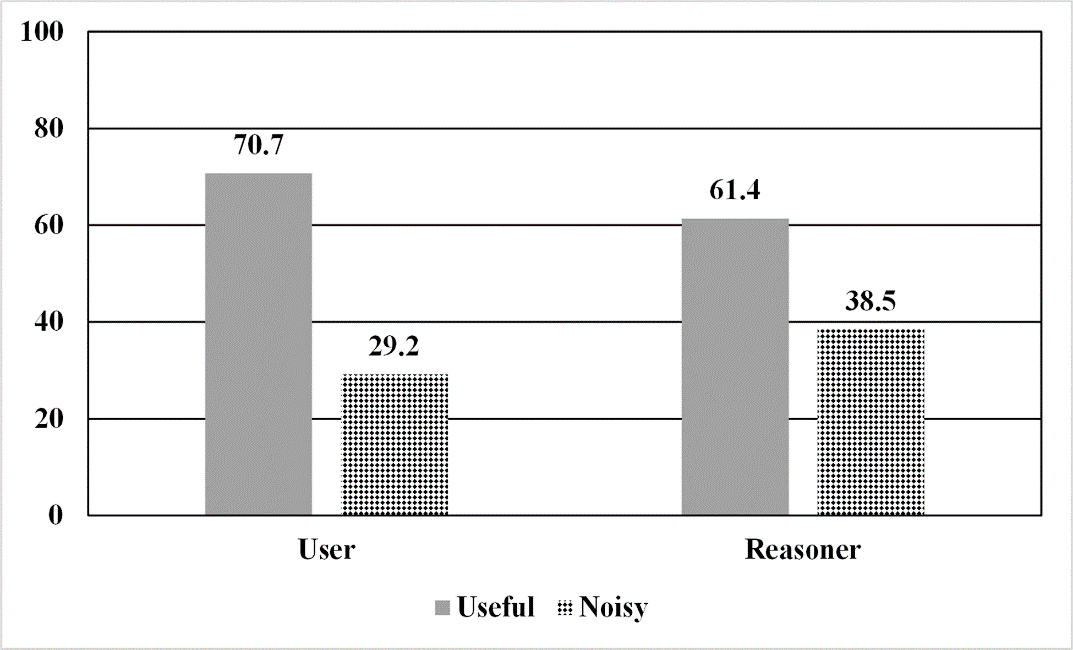}
    \caption{ Percentage of hyperlink types as perceived by the user and the reasoned}
    \label{fig:life}
\end{figure}

Figure 14 shows that the reasoner and the user are able to detect useful hyperlinks in 61.4 and 70.7 percent of the times. With respect to noisy hyperlinks, the values are 38.5 and 29.2, respectively. Based on these results, it is clear that the reasoner is able to achieve the results of an expert user. 

\subsubsection{ Scalability }
It is important that the final data collection contain records which can be analyzed through semantic and relatedness approaches. In other words, the data collection must have records which belong to the classes of the DBpedia ontology. The reasoner requires a certain amount of time to discover new relations, classes, and properties based on the DBpedia ontology and determine the type (i.e. noisy or useful) of the hyperlinks in the data collection. In the remainder of this section we discuss scalability of the reasoner as well as the entire process. Scalability tests were performed a computer system with an Intel Core i5 2450M (2.50 GHz) with 4.00 GB of RAM and running a 64-bit operating system. In Figure 15, the amount time required to run the semantic and relatednessreasoner for 500, 1000, 1500, and 1946 hyperlinks can be seen. 
   \begin{figure}
    \centering
    \includegraphics[width=9cm]{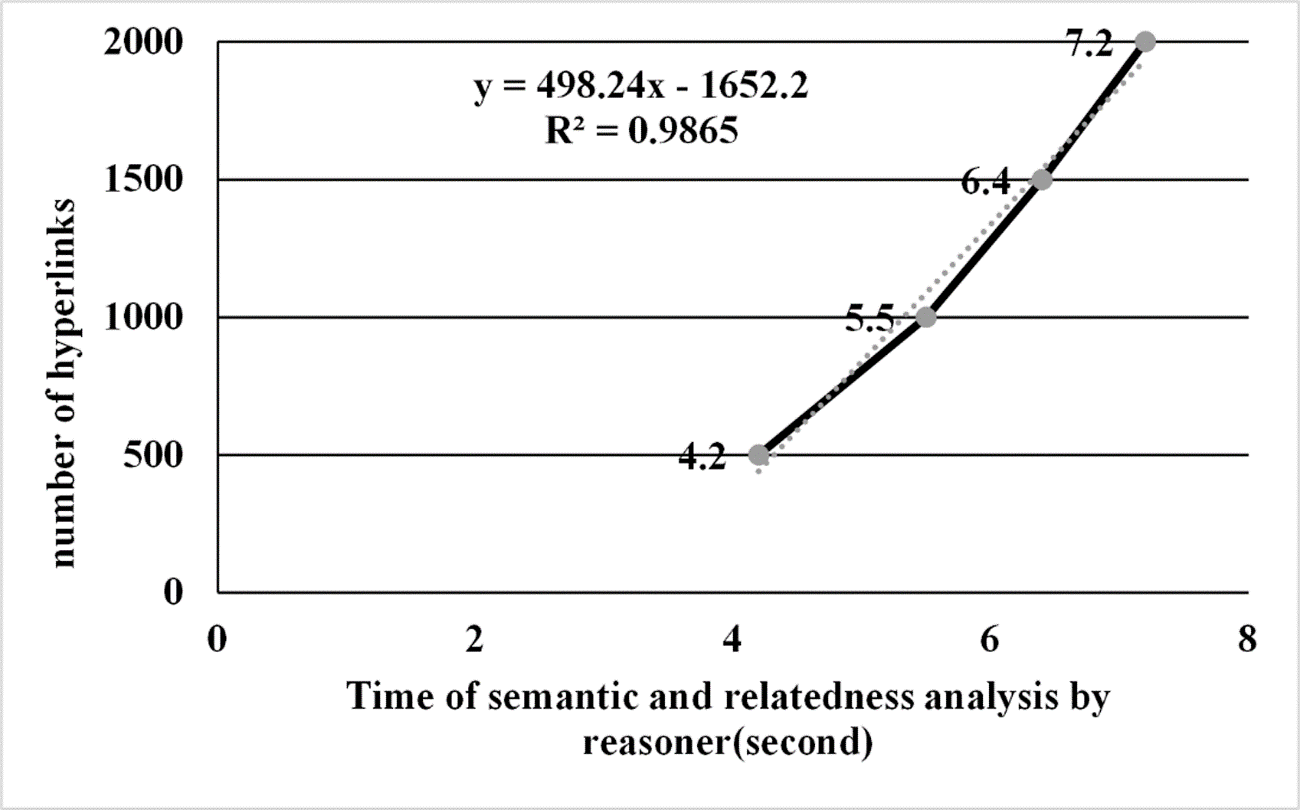}
    \caption{Scalability of the reasoner for different numbers of hyperlinks}
    \label{fig:life}
\end{figure}

The variables y and R denote the number of hyperlinks and the accuracy of the equation, respectively. During the initial two to three seconds, the reasoner is being activated. In Figure 15, it is assumed that the final dataset is available. However, in Figure 16, scalability of the entire process from topic identification to semantic and relatedness analysis is considered for the same cases. 
   \begin{figure}
    \centering
    \includegraphics[width=9cm]{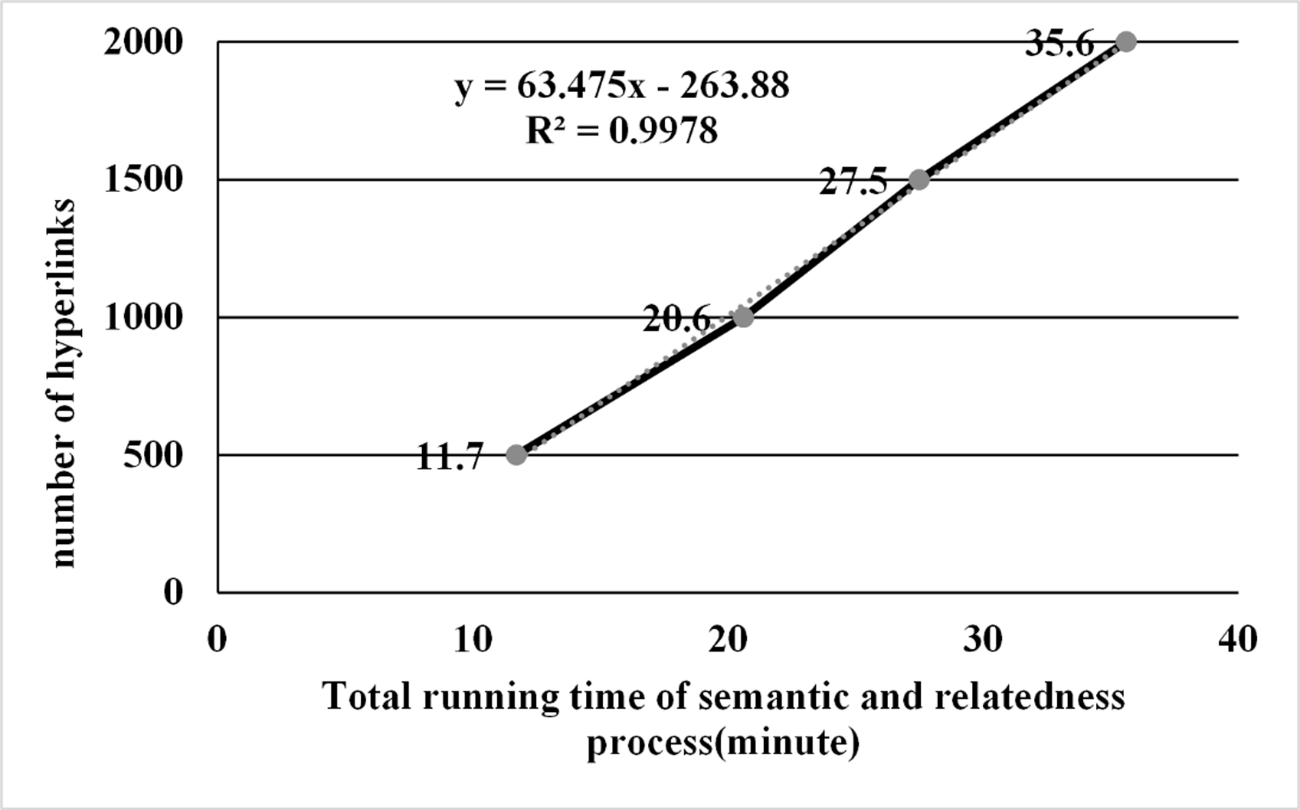}
    \caption{ Scalability of the entire system for different numbers of hyperlinks}
    \label{fig:life}
\end{figure}

Our investigations revealed that the matching stage is the most time-consuming step of the process.

\section{Discussion and Conclusion }
Hyperlinks are a type of noisy data in the structure of the web, which negatively impact the efficiency of many information retrieval algorithms. Nearly all of these algorithms focus on the string or graph structure of hyperlinks. Therefore, these approaches incorrectly remove certain useful hyperlinks and are not able to detect noisy hyperlinks in certain cases. In this paper, we consider semantic and relational structures at both the page and site levels while semantic web technologies such as ontologies and reasoners are used to eliminate noisy hyperlinks. In doing so, a dataset of hyperlinks is created in a separate process. The dataset is then analyzed with respect to both semantics and relatedness. As a result of the analysis, noisy and useful hyperlinks are distinguished.
The proposed system takes the constructed dataset as input. Each row of the dataset consists of the class mapped from the hyperlink context topic of the source page, the class mapped from the topic of the target page, a field indicating the noisy or useful nature of the hyperlink from the user’s perspective, and the domain name of the source page. Then the results are compared to those of the user to demonstrate the extent to which each semantic or relational property in the ontology contributes to a hyperlink being identified as either noisy or useful. Furthermore, the categories of queries which lead users to noisy hyperlinks and the domains with the highest number of noisy hyperlinks can be found. Our experiments proved the accuracy, ability, and scalability of semantic web technologies in eliminating noisy hyperlinks. Directions for future works include the following: \\
1.	Combining the DBpedia ontology with another ontology to cover a larger domain at the T-Box level concepts. \\
2.	Using available datasets on linked data to cover A-Box level concepts. \\
3.	Extending the noise detection operation to include hyperlinks pointing to images, videos, and audio files. \\
4.	Applying various algorithms to match topics to ontology classes. \\
5.	Using semantic tools to identify the topic of hyperlinks and target pages via semantic properties in pages that are constructed using Web 3.0 techniques.\\

\bibliographystyle{IEEEtran}
\bibliography{References}

\end{document}